\documentclass[prb,showpacs,amsmath,amssymb,superscriptaddress,twocolumn,floatfix]{revtex4-1}

\usepackage{graphicx}
\usepackage{color}
\usepackage{amsmath}
\usepackage{hyperref}
\hypersetup{breaklinks=true}

\begin{document}

\title{Non-Relativistic Anisotropic Magnetoresistance with Collinear and Non-Collinear Magnetic Order}

\author{P. Ritzinger}

\author{O. Sedl\'a\v cek}
\author{J. \v Zelezn\'y}

\author{K. V\'yborn\'y}
\affiliation{Institute of Physics, ASCR, $v.~v.~i.$, 
Cukrovarnick\'a 10, CZ-16253 Praha 6, Czech Republic}

\date{Jun27, 2025}

\begin{abstract} 
Anisotropic magnetoresistance (AMR) arises from symmetry lowering of the conductivity tensor induced by magnetic order.
In simple ferromagnets, AMR is a relativistic effect, relying on spin-orbit interaction (SOC). Here, we demonstrate that a comparable symmetry lowering can also occur in a non-relativistic limit. Using tight-binding models, density functional theory calculations, and Boltzmann transport theory, we investigate systems with multiple magnetic sublattices, including both collinear and non-collinear antiferromagnets, as well as ferrimagnetic configurations. We show that AMR and related anisotropies can emerge purely from magnetic order, without the need for SOC, and may reach significant magnitudes. The findings are supported by case studies on toy-model lattices and real materials such as MnN, Mn$_3$Sn, and are further interpreted using a symmetry analysis based on Neumann’s principle. 
Material candidates that exhibit non-relativistic anisotropic magnetoresistance are identified
by symmetry analysis applied to entries in the MAGNDATA database.
\end{abstract}

\maketitle

\section{Introduction}

Anisotropic magnetoresistance (AMR), first observed by William Thomson in 1857~\cite{Thomson:1857}, describes the dependence of resistivity in ferromagnetic materials such as cobalt and nickel on the direction of magnetization. Since its discovery, AMR has remained an active subject of research~\cite{Ritzinger:2023}. In its prototypical form, AMR appears in polycrystalline ferromagnets (FMs) as a two-fold (that is 180$^\circ$--periodic) angular dependence of resistivity $\rho$ on the angle $\varphi$ between the current and the magnetization direction~\cite{Alagoz:2015}:
\begin{equation}
	\frac{\Delta \rho}{\rho} \propto \cos 2\varphi.
	\label{eq_ncollAMR}
\end{equation}
This relation implies that, for example, if the magnetization in an otherwise isotropic medium ($\sigma_{xx} = \sigma_{yy}$ above the magnetic ordering temperature) is oriented along the $x$-direction, the conductivity tensor $\boldsymbol{\sigma} = \boldsymbol{\rho}^{-1}$ becomes anisotropic with $\sigma_{xx} \neq \sigma_{yy}$.

The conventional understanding of AMR~\cite{Vitayaya:2024}, in line with Eq.~\ref{eq_ncollAMR}, attributes the effect to spin-orbit coupling (SOC), which couples the spin of electrons to the crystal lattice. However, this description is not exhaustive. In this work, we discuss similar symmetry-lowering effects in the conductivity tensor that can also arise in the absence of SOC, provided the magnetic order involves multiple magnetic sublattices (MSLs), including both collinear and non-collinear magnetically ordered systems.

\subsection{Definition of AMR}

More generally, AMR can be defined as the change in symmetry of the conductivity tensor $\sigma_{ij}$ due to magnetic ordering. This definition includes the standard expression in Eq.~\ref{eq_ncollAMR}, but excludes conductivity anisotropies
due to the direct action of the magnetic field, such as orbital effects in non-magnetic systems (e.g., positive magnetoresistance in metals which may be anisotropic~\cite{Zhang:2019}),
or surface states in topological insulators~\cite{Ritzinger:2023}. Historically, AMR has also been referred to by various terms—such as spontaneous magnetoresistance anisotropy (SMA)—though these are not always consistently defined throughout the literature~\cite{Ritzinger:2023}. The above definition thus provides a unifying framework for these differing terminologies.

The conventional understanding of AMR encompasses two distinct phenomena: 1) the anisotropy of the conductivity induced by magnetic order and 2) the variation of the conductivity when the magnetic order is rotated. In many commonly studied cases, these two manifestations are effectively equivalent, and the distinction between them is often overlooked. This equivalence can be illustrated using a simple example: an FM on a square lattice. In the absence of magnetic order, the conductivity is isotropic. Once ferromagnetism is present, the conductivity becomes anisotropic:

\begin{equation}
	\sigma(\mathbf{M} \parallel x) =
	\begin{pmatrix}
		\sigma & 0 \\
		0 & \sigma'
	\end{pmatrix},
	\quad
	\sigma(\mathbf{M} \parallel y) =
	\begin{pmatrix}
		\sigma' & 0 \\
		0 & \sigma
	\end{pmatrix}
\end{equation}

The AMR is then defined as $\frac{\sigma-\sigma'}{\sigma}$, which can be measured both by measuring the conductivity along different directions for fixed magnetization:

\begin{equation}
\text{AMR} = \frac{\sigma_{xx}(\mathbf{M}  \parallel x) - \sigma_{yy}(\mathbf{M}  \parallel x)}{\sigma_{xx}(\mathbf{M}  \parallel x)} = \frac{\sigma - \sigma'}{\sigma}
\label{AMR_Eq3}
\end{equation}
as well as by rotating the magnetization and measuring conductivity along a fixed direction:

\begin{equation}
\text{AMR} = \frac{\sigma_{xx}(\mathbf{M}  \parallel x) - \sigma_{xx}(\mathbf{M}  \parallel y)}{\sigma_{xx}(\mathbf{M}  \parallel x)} = \frac{\sigma - \sigma'}{\sigma}
\label{AMR_Eq4}
\end{equation}
where the second approach is typically used in experiments, as it is 
easier to measure accurately in practice.

Although the two approaches are equivalent in simple cases, they differ in general. Without SOC, the spin and orbital degrees of freedom are decoupled. As a consequence, the non-magnetic lattice is symmetric under any pure spin rotation. In magnetic systems, a rigid rotation of all magnetic moments corresponds to such a pure spin rotation, and since conductivity is invariant under spin rotations, rotating the magnetic order (as a whole) cannot modify the conductivity in the absence of SOC.
Therefore, the second form of AMR, involving rotation of magnetic order, does not exist without SOC.
However, we demonstrate in this work that the first form---anisotropy induced by magnetic order itself---can persist without SOC in systems featuring multiple magnetic sublattices. 
Moreover, we consider additional cases involving more complex manipulations of magnetic order (e.g., non-rigid tilting) and scattering from ferromagnetically aligned impurities in non-collinear systems, which can also lead to AMR in the non-relativistic limit.

Throughout this study, we focus exclusively on longitudinal AMR, i.e., anisotropy in the diagonal components of the conductivity tensor. We note that AMR can also appear in off-diagonal components,\cite{Badura:2025_a} often referred to as the planar Hall effect~\cite{Ritzinger:2023}. Furthermore, we restrict our analysis to the $\cal{T}$-even component of the conductivity, which corresponds to the symmetric part of the conductivity tensor. The $\cal{T}$-odd part, associated with the anomalous Hall effect (AHE), is not studied here. While AHE can also occur without SOC, it requires non-coplanar magnetic systems~\cite{Zhang:2018}, a scenario typically described as the topological Hall effect.

\subsection{Categories of AMR}

Beyond its mode of realization, AMR can be categorized based on its microscopic origin:

\paragraph{Intrinsic vs. Extrinsic.} Intrinsic AMR originates from symmetry-breaking effects that are scattering-independent, whereas extrinsic AMR arises from spin-dependent scattering. Historically, the extrinsic mechanism has received more attention, and only few works have recognized scattering-independent contributions to AMR~\cite{Kato:2008, Velev:2005, Zeng:2020, Kato:2007, Nadvornik:2021_a, Park:2021} so far. 
These two contributions can be experimentally distinguished using frequency-dependent measurements~\cite{Nadvornik:2021_a, Park:2021}: the extrinsic contribution typically scales with $1/\omega$, while the intrinsic one is frequency-independent. 
In the context of other transport phenomena such as the AHE or spin Hall effect (SHE), intrinsic mechanisms are well-established and linked to the Berry curvature~\cite{Nagaosa:2010, Zhang:2017}. In contrast, the origin of intrinsic AMR is usually attributed to Fermi surface anisotropy, which we address in Sec.~\ref{sec_intrinsic}. However, a recent study~\cite{Dong:2025_a} suggests that Berry curvature may also play a role in intrinsic AMR. An extreme form of intrinsic AMR can be caused by a metal-insulator transition (MIT) in EuTe$_2$, which exhibits large AMR values (up to 40,000\%)~\cite{Yang:2021} when measured along different crystallographic directions under applied magnetic fields. Here, the Fermi surface (FS) does not become anisotropic, but rather vanishes entirely in the insulating phase, leading to a dramatic conductivity anisotropy.

\paragraph{Non-Crystalline vs. Crystalline.} The AMR signal in polycrystalline samples follows the $\cos 2\varphi$ dependence of Eq.~\ref{eq_ncollAMR} and is called non-crystalline AMR. In contrast, high-quality single crystals can exhibit more complex angular dependencies. For example, four-fold symmetry has been reported in Ni~\cite{Doring:1938}, Co$_2$MnGa~\cite{Sato:2019,Ritzinger:2021}, and (Ga,Mn)As~\cite{DeRanieri:2008}; six-fold symmetries in hexagonal MnTe~\cite{Kriegner:2017, Gonzalez-Betancourt:2024}; and sometimes even higher-frequency periodicity in certain cases~\cite{DeRanieri:2008,Gonzalez-Betancourt:2024, NamHai:2012}. Although such features have been known for decades, they are occasionally misinterpreted as manifestations of magnetocrystalline anisotropy~\cite{Ritzinger:2023} or highlighted as recent discoveries, though related concepts have been explored in earlier works~\cite{Dong:2023}. Typically, higher angular harmonics are analyzed in systems with a single spin axis (SSA). However, this analysis can be straightforwardly extended to systems with multiple magnetic sublattices (MSLs), as demonstrated in Appendix~\ref{apx_phenomodel}.

This section provides only a brief overview of the rich landscape of AMR phenomena. For a more comprehensive treatment, we refer the reader to Ref.~\onlinecite{Ritzinger:2023}.

\subsection{Beyond Spin-Orbit Coupling}

This study builds on earlier findings (see Sec. 4.2.3 of Ref.~\onlinecite{Ritzinger:2023}) 
of spin-order induced anisotropies which fall, in the present context, into the scope of
non-relativistic AMR. Recently, various effects that typically originate from spin-orbit coupling have been shown to exist in magnetic systems even with no SOC.
For instance, in the non-collinear antiferromagnet Mn$_3$Sn, both a strong SHE~\cite{Zhou:2020, Zhang:2017} and local (sublattice-projected) Edelstein effect~\cite{Gonzalez-Hernandez:2024} were found to prevail even in the absence of SOC~\cite{Manna:2018, Gonzalez-Hernandez:2024}, indicating that complex magnetic order can mimic relativistic effects. Non-relativistic effects have also been explored within the emerging framework of altermagnetism~\cite{Jungwirth:2024, Smejkal:2022}. Non-relativistic parity-breaking of Fermi surfaces (by magnetic order) leading also to conductivity anisotropies was investigated in a separate study~\cite{BirkHellens:2023}, which focuses on the symmetry-based characterization of magnetic phases exhibiting such features.
The AMR alone, however, does not require parity breaking, and simpler scenarios can occur.

\subsection{Organization}

In this work, we explore how AMR can arise in the absence of SOC, considering both intrinsic and extrinsic mechanisms:~\cite{Vyborny:2009} \\

\textbf{Intrinsic AMR}: We identify two distinct mechanisms for magnetic-order-induced anisotropy in the conductivity tensor that emerge without spin-orbit coupling (SOC), both assuming isotropic scattering:
	
\begin{enumerate}
	\item \emph{Spontaneous anisotropy}: Intrinsic anisotropy arising from magnetic ordering itself, as observed in collinear antiferromagnets like MnN. \textit{Ab initio} calculations indicate this effect can be significant, though it typically offers limited tunability via external stimuli (e.g., magnetic fields).
	
	\item \emph{Field-induced anisotropy}: Manipulation of magnetic order via applied fields or strain, where we consider tilting of the moments rather than a rotation. We explore this in (i) minimal models (kagome/triangular lattices), and (ii) the non-collinear antiferromagnet Mn$_3$Sn.
\end{enumerate}
	
\textbf{Extrinsic AMR:} We also show that even in systems where the magnetic order does not induce any non-relativistic intrinsic AMR, extrinsic AMR can exist when a specific type of scattering is considered. We assume scattering on magnetic impurities, such that typically unordered impurities are only weakly coupled to the lattice and can be aligned by a weak external magnetic field, which does not overcome the exchange interactions of the host material.
Our motivation here are weakly coupled Mn impurities in antiferromagnetic salts and alloys 
which we will discuss in detail in Sec.~\ref{sec_I_mat}.

This paper is organized as follows: In Sec.~\ref{sec_modelling}, we
will introduce the theoretical background and the methodology used in this work. In Sec.~\ref{sec_intrinsic}, we will discuss the intrinsic AMR, followed by the discussion of extrinsic AMR in Sec.~\ref{sec_extrinsic}. The summary follows in the final Sec.~\ref{sec_Sum}. 

\section{Methods}
\label{sec_modelling}

\subsection{Formalism}

Except for Sec.~\ref{sec_I_MnN}, we rely on a simple tight-binding model which only consists of a hopping and an exchange term~\cite{Gonzalez-Hernandez:2024}:
\begin{equation}
	H = -\sum_{i, j}\sum_\alpha t_{ij} {\hat{c}_i^{\alpha\dagger}} \hat{c}^\alpha_j + J \sum_{i} \sum_{\alpha, \beta} (\mathbf{\sigma} \cdot \hat{m}_i)_{\alpha \beta} {\hat{c}_i^{\alpha\dagger}} \hat{c}^\beta_i 
	\label{eq_sdmodel}
\end{equation} 
where $t_{ij}$ is the hopping parameter from site $i$ to $j$, $\alpha$ and $\beta$ are the spin indices, ${\hat{c}_i^{\alpha}}(^\dagger)$ is an annihilation (creation) operator at site $i$ with spin $\alpha$, $J$ is the Heisenberg exchange constant, $\mathbf{\sigma}$ the vector of the Pauli spin matrices and $\hat{m}_i$ the magnetization direction unit vector at site $i$.

The conductivity is then calculated using the Boltzmann equation~\cite{Vyborny:2009}. 

\begin{multline}
	\sigma_{ij} = e^2 \sum_n  \int_ {1st BZ} \frac{d^3k}{(2\pi)^3} \delta(E_n(\mathbf{k}) - E_F) \frac{1}{\hbar \Gamma_{n, \mathbf{k}}} \times \\ v_{n,i}(\mathbf{k}) v_{n,j}(\mathbf{k})
	\label{eq_Boltzmann_1}
\end{multline}
%
where $e$ is the elementary charge, $E_n(\mathbf{k})$ is the $k$-dependent eigenenergy of the $n$th-band, $E_F$ is the Fermi energy, $\Gamma_{n, \mathbf{k}}$ is the scattering rate and $v_{n,i}$ is the $i$-th component of the Fermi velocity in the $n$-th band. The delta distribution evaluates the integral over the first Brillouin zone (1st BZ) at the Fermi surface. The Fermi velocity is calculated by:
\begin{equation}
	v_{n, i} = \frac{1}{\hbar} \frac{\partial E_n(\mathbf{k})}{\partial k_i}
\end{equation}
In case of the intrinsic AMR (Sec.~\ref{sec_I_Kagome} and \ref{sec_I_mat}), the scattering rate is obtained by choosing the relaxation-time approximation (RTA)~\cite{Vyborny:2009_a}
where the relaxation time $\tau\propto 1/\Gamma_{n, \mathbf{k}}$ is constant and thus isotropic. This means that the anisotropy can only enter through the Fermi velocity contribution (or anisotropic plasma frequencies\cite{Ambrosch-Draxl:2006} $\omega^p$). 
We can simplify Eq.~\ref{eq_Boltzmann_1} to:
\begin{equation}
	\sigma_{ii} \propto \int_ {FS} \sum_n   d\mathbf{k}  v^2_{n,i}(\mathbf{k})
	\propto (\omega_{ii}^p)^2
	\label{eq_Boltzmann_2}
\end{equation}
where $\mathbf{k}$ is the wave vector running over the Fermi surface.
We made use of the facts that we only consider the
longitudinal conductivity $\sigma_{ii}$ in two-dimensional systems,
the delta distribution evaluates the integral over the first Brillouin
zone at the FS, and the scattering rate is obtained by
RTA. The anisotropy $\sigma_{xx} \neq \sigma_{yy}$ arises if the integral of $\sum_n
v^2_{n,x}$ and $\sum_n v^2_{n,y}$ over the Fermi surface are not the
same. This generally requires an anisotropic FS that breaks crystal symmetries. A spherical Fermi surface or one that follows the full crystal symmetry (e.g., hexagonal in a hexagonal lattice) does not lead to anisotropic conductivity, in accordance with Neumann’s principle.

For extrinsic AMR (Sec.~\ref{sec_extrinsic}), we return to Eq.~\ref{eq_Boltzmann_1}, and determine the scattering rate by using Fermi's Golden Rule:
\begin{multline}
	\Gamma_{n, \mathbf{k}} = \frac{2 \pi}{\hbar} N_{scat} \sum_{n'}  \int_ {1st BZ} \frac{d^3k'}{(2\pi)^3} \delta(E_{n'}(\mathbf{k'}) - E_n(\mathbf{k})) \times \\ |M^{\mathbf{k}\mathbf{k'}}_{nn'} |^2 (1 - \cos \theta_{vv'})
	\label{eq_FermiGoldenRule_1}
\end{multline}
where $N_{scat}$ is the volume density of the scatterers, $M^{\mathbf{k}\mathbf{k'}}_{nn'}$ is the transition matrix element and $\cos \theta_{vv'} = \frac{\mathbf{v}_n (\mathbf{k})}{|\mathbf{v}_n (\mathbf{k})|}\frac{\mathbf{v}_n' (\mathbf{k'})}{|\mathbf{v}_n' (\mathbf{k'})|}$. The transition matrix element is calculated by:

\begin{equation}
	M^{\mathbf{k}\mathbf{k'}}_{nn'} = \langle \psi_{n, \mathbf{k}}|\hat{M}|\psi_{n', \mathbf{k'}} \rangle
	\label{eq_transmatrix}
\end{equation}
where $\psi_{n, \mathbf{k}}$ is the wave function for the eigenenergy value $E_n(\mathbf{k})$~\cite{Vyborny:2009}
and $\hat{M}$ describes the particular form of impurity.

\subsection{Symmetry Analysis}
\label{sec_symmAna}

We will analyze the real-space symmetries of various magnetic configurations—both for the toy models and Mn$_3$Sn—using the open-source code \textit{Symmetr}~\cite{Symmetr}. The software identifies the non-relativistic symmetry group of each magnetic configuration and returns the corresponding generator matrices and the symmetry-restricted form of response tensors such as the conductivity tensor. A more detailed discussion about the non-relativistic symmetry analysis is provided in Ref.~\onlinecite{Gonzalez-Hernandez:2024}. To understand the origin of anisotropic magnetoresistance (AMR), we compare the symmetries of configurations that exhibit isotropic conductivity with those that show AMR. The key idea is to identify which symmetries must be broken to enable AMR.

At the core of this analysis lies Neumann’s principle, which states that a tensor representing a macroscopic physical property of a crystal must be invariant under the symmetry operations of that crystal.\cite{Ritzinger:2021}

\section{Intrinsic AMR}
\label{sec_intrinsic}

In this section, we discuss the mechanism sketched in Fig. 1a of Ref.~\onlinecite{Vyborny:2009} where the anisotropy is not related to scattering processes. Unlike in that reference, the intrinsic~\cite{Nadvornik:2021_a} AMR here is not due to SOC, but instead arises from magnetic order. Subsection~\ref{sec_I_MnN} focuses on MnN as an example of spontaneous AMR in a collinear antiferromagnet and further materials are discussed in Sec.~\ref{sec_Table}. In subsection~\ref{sec_I_Kagome}, we investigate various magnetic configurations—both collinear and non-collinear, compensated and non-compensated—on kagome and triangular lattices. Using a tight-binding model and the Boltzmann formalism in the RTA, we analyze Fermi surface anisotropy and conductivity. We also examine the real-space symmetries underlying these models. In subsection~\ref{sec_I_mat}, we apply the same framework to Mn$_3$Sn as an example of currently topical AFM.

\subsection{MnN: Example of AMR as a Spontaneous Effect}
\label{sec_I_MnN}

\begin{figure}
	\centering
	\includegraphics[width=1\linewidth]{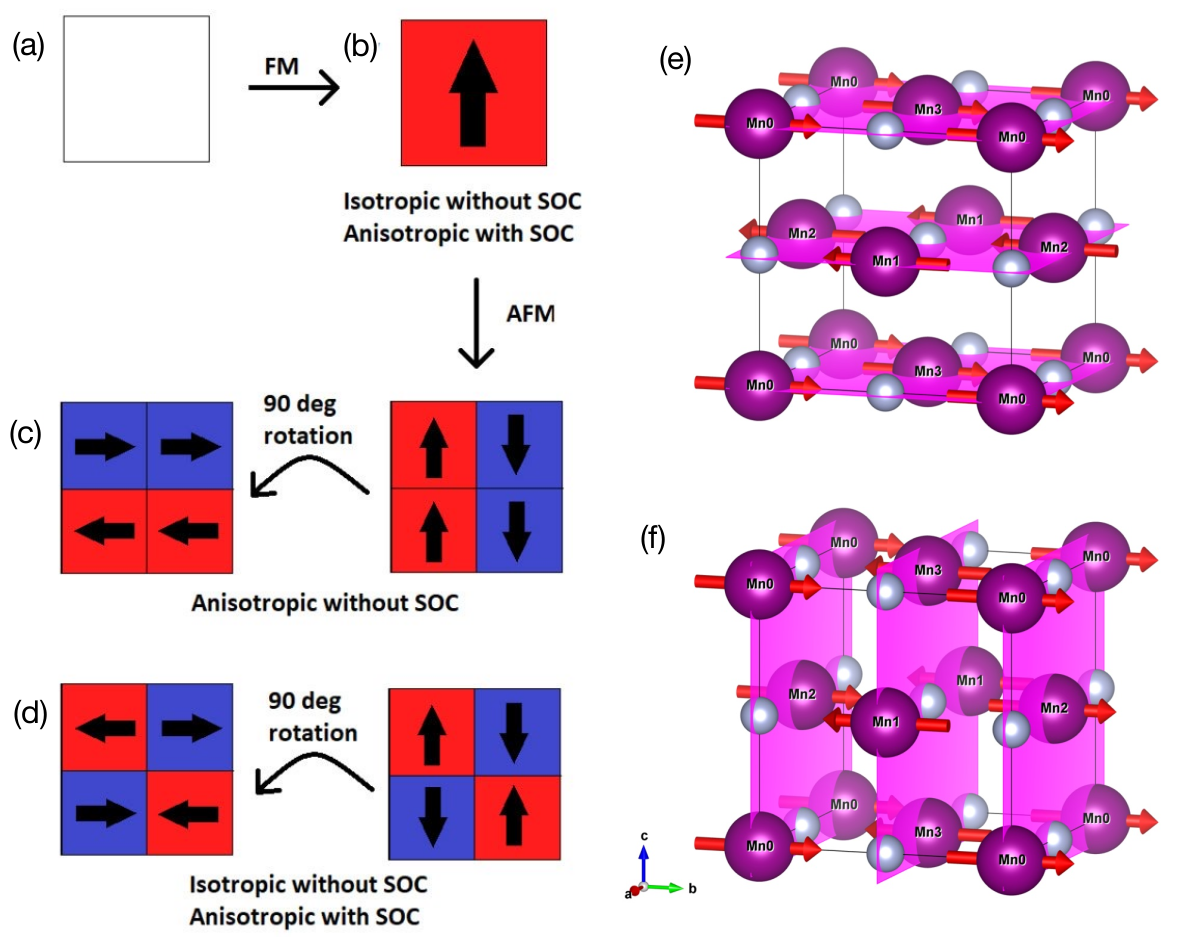}
	\caption{(a) A square lattice without magnetic order exhibits isotropic conductivity $\sigma_{xx} = \sigma_{yy}$. (b) FM order allows for AMR $\sigma_{xx} \neq \sigma_{yy}$ only in the presence of SOC. (c,d) Introducing multiple MSL (red and blue) allows for AMR in a collinear AFM even in the absence of SOC if the $90^\circ$ real-space rotation symmetry is broken by the magnetic order: (c) Symmetry-breaking due to the rotation of the FM planes. (d) The FM planes remain invariant under real-space rotation. The conductivity will remain isotropic due to the linearity of the conductivity tensor. (e,f) Crystalline structure with two different types of magnetic order in MnN. In (f), Mn2 and Mn3 are flipped in comparison to (e) leading to  90$^\circ$ rotation of the ferromagnetic planes (highlighted in violet), allowing for non-relativistic AMR.
    }
	\label{fig:mnnsketch}
\end{figure}

AMR as a spontaneous effect
is illustrated in Fig.~\ref{fig:mnnsketch}: In a cubic crystal without magnetic order, the conductivity remains isotropic $\sigma_{xx}=\sigma_{yy}=\sigma_{zz}$. A FM with a single MSL breaks the symmetry and leads to anisotropic conductivity in the presence of SOC. However, in the non-relativistic limit, the conductivity remains isotropic. This is because, without SOC, spin and real space are decoupled, and the four-fold rotational symmetry of the cubic crystal will not be broken by the FM order. More generally, with SOC, 
rotational symmetry operations involve both real-space and spin-space components. The same holds for mirrors, except that a mirror is described by improper rotation in real space and by corresponding proper rotation in spin space.
Without SOC, however, the real-space and spin-space components of symmetry operations are decoupled. 
In systems with a single MSL, all crystallographic symmetries of the nonmagnetic lattice (excluding time-reversal) remain valid, as the spin component can be considered trivial (i.e., no transformation in the spin space). 
Since we only consider the $\cal{T}$-even part of the conductivity, which does not depend on spin-space transformations, FM order without SOC does not change the symmetry of the conductivity tensor.

However, in systems with multiple MSLs, symmetry operations can map one sublattice onto another. In this case, even in the absence of SOC, the magnetic order can break crystallographic symmetries and thereby alter the symmetry of the conductivity tensor. %
This is illustrated in Fig.~\ref{fig:mnnsketch}c, which shows an A-type AFM order, consisting of FM planes. This configuration breaks the cubic symmetry of the crystal because a $90^\circ$ rotation rotates the direction of the FM planes. Since this direction does not depend on the direction of the SSA, no spin rotation could be combined with the $90^\circ$ rotation to make it a symmetry. In contrast, an AFM order shown in Fig.~\ref{fig:mnnsketch}d in which the ferromagnetic planes are oriented along the diagonal direction, preserves the four-fold symmetry and does not induce any anisotropy in the conductivity tensor. 

The type of magnetic order, shown in Fig.~\ref{fig:mnnsketch}c, occurs in MnN, which has a rock-salt structure (see ~\ref{fig:mnnsketch}e): Without magnetic order, the conductivity remains isotropic. However, the cation (manganese) magnetic moments prefer an A-type AFM order. When choosing the direction of the ferromagnetic planes, e.g, $xy$-planes, anisotropy arises---in this case, $\sigma_{zz}$ is different from $\sigma_{xx}=\sigma_{yy}$~\cite{Granville:2005}. Our LAPW~\cite{Blaha:1990} calculations based on density functional theory for a perfectly cubic structure show that $\hbar\omega^p_{xx}=5.71$~eV and $\hbar\omega^p_{zz}=5.29$~eV so that, assuming isotropic scattering, non-zero AMR according to Eq. \ref{eq_Boltzmann_2} exists, with $\sigma_{xx}/\sigma_{zz}-1\approx 17$~\%. In comparison, for calculations using GGA with SOC, we obtained $\hbar\omega^p_{xx}=5.87$~eV and $\hbar\omega^p_{zz}=5.23$, leading to similar anisotropies and thus, indicating that the AMR is mainly of non-relativistic nature. 
Although magnetic ordering leads to a slight distortion of the lattice, this has only a minor influence on the transport anisotropy. For example, when neglecting SOC and choosing $a/c=0.4256/0.4189$~nm, the plasma frequencies change to $\hbar\omega^p_{xx} = 5.88$~eV and $\hbar\omega^p_{zz} = 5.10$~eV.

\subsection{Further Material Candidates}
\label{sec_Table}

Although the occurrence of non-relativistic AMR in antiferromagnetic MnN was discussed in detail in the previous section, this material represents only one example within a broader class of candidates. By applying the symmetry analysis introduced in Sec.~\ref{sec_symmAna} to a wider range of materials, we identified 280 potential candidates, as detailed in the Supplementary Material. Prominent examples include Mn$_5$Si$_3$, which exhibits topological Hall effect in its low-temperature non-coplanar phase~\cite{Surgers:2014}
the non-collinear compound MnPtGa, which exhibits an AHE~\cite{Ibarra:2022} alongside a pronounced magnetocaloric effect~\cite{Cooley:2020}; and Sr$_2$IrO$_4$, a canted antiferromagnet hosting a novel spin-orbit entangled $J_\text{eff}=1/2$ state studied for its similarity to cuprate superconductors~\cite {Yan:2015}. The crystal and magnetic structures used in our analysis were obtained from the MAGNDATA database~\cite{Gallego:2016_a, Gallego:2016_b}

\subsection{Intrinsic AMR due to Manipulation of the Magnetic Order}
\label{sec_I_Kagome}

\begin{figure}[b]
	\centering
	\includegraphics[width=1\linewidth]{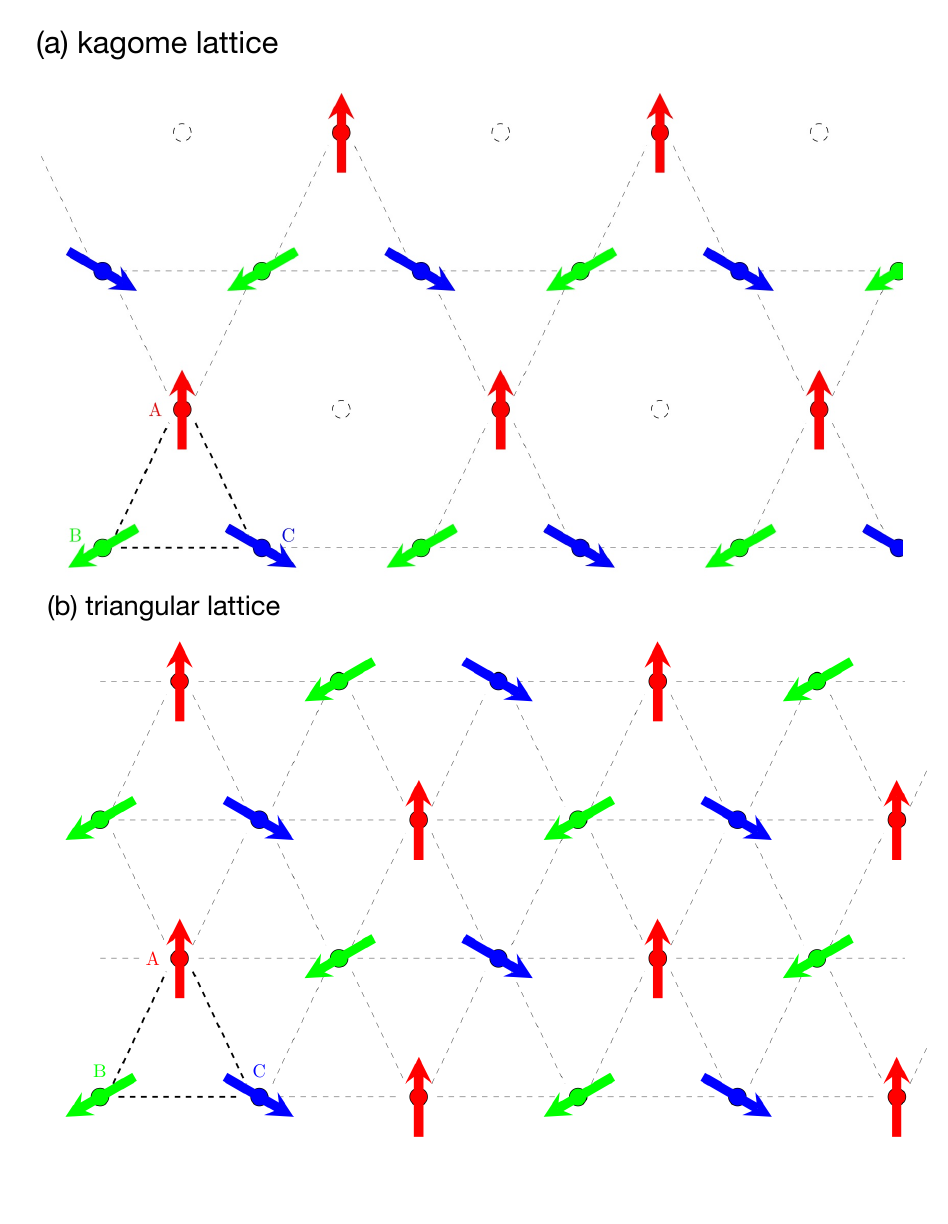}
	\caption{Schematic illustration of (a) the kagome lattice and (b) the triangular lattice with non-collinear magnetic order. The three magnetic moments in the magnetic unit cell are shown in red (A), green (B), and blue (C). The net magnetic moment is zero. While both configurations share the same magnetic unit cell, the kagome lattice features a regular vacancy (dashed circle), distinguishing it from the triangular lattice. In practice, the vacancy can be filled with a non-magnetic atom.}
	\label{fig:kagome_triangular}
\end{figure}

\begin{figure}[h!]
	\centering
	\includegraphics[width=1\linewidth]{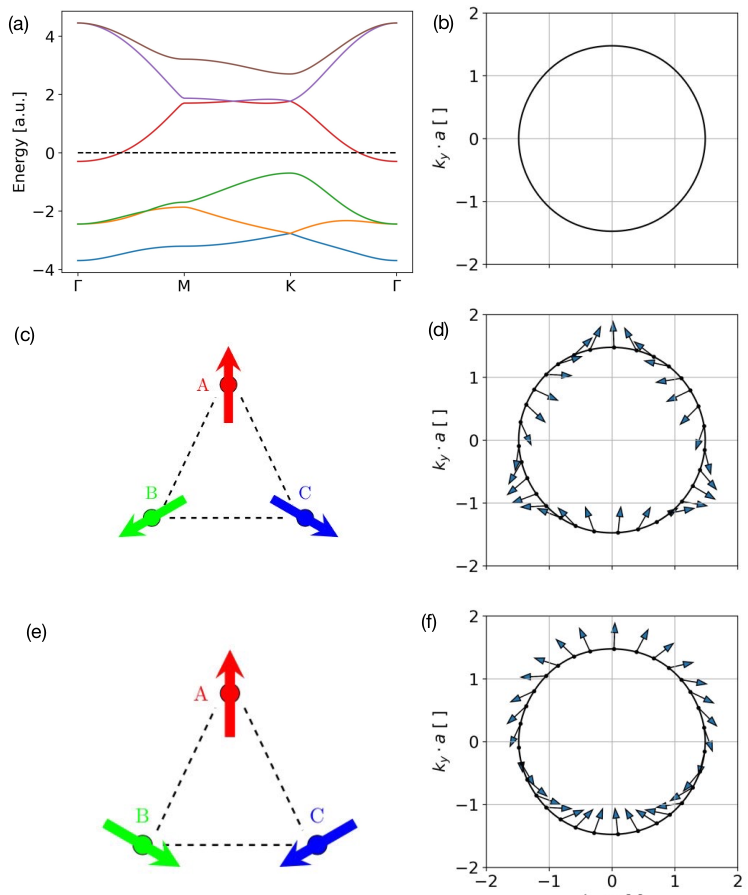}
	\caption{Results for two compensated magnetic configuration on the kagome lattice. (a) The band structure is spin-split. Permutation, which corresponds to a $180^\circ$ global spin rotation around the \textit{y}-axis, does not change the band structure. (b) The Fermi surface at $E_F = 0$ is isotropic and shows no intrinsic AMR. Note that a circular FS is not the only shape that conserves the isotropy of conductivity 
    (c), (e) The magnetic unit cells of two compensated magnetic configurations and (d), (f) their respective k-space spin texture at the same Fermi level. While the band structure and Fermi surface are unaffected by permutations within the magnetic unit cell, the spin texture changes. The spin texture lies entirely in the \textit{xy}-plane. The $z$-component (not shown) of the spin texture is zero.}
	\label{fig:totalkagomephase1a}
\end{figure}

 In this section, we study how intrinsic AMR can be induced by manipulating the magnetic order, which can be achieved, for example, by means of an applied magnetic field or strain. We consider a triangular magnetic order illustrated in Fig.~\ref{fig:kagome_triangular}, as this type of order is found in many materials~\cite{Siddiqui:2020}. It arises from antiferromagnetic exchange of three spins to avoid frustration. Contrary to MnN, where the magnetic order itself breaks the real-space rotational symmetry, non-collinear magnetic order on a kagome and triangular lattice preserves the symmetry of the conductivity tensor as shown in Fig.~\ref{fig:totalkagomephase1a}(b). As discussed earlier, a global rotation of all magnetic moments corresponds to a pure spin rotation, which does not alter the conductivity. This is reflected in the unchanged band structure shown in Fig.~\ref{fig:totalkagomephase1a}. Although the spin texture rotates under such transformations, the band structure remains unaffected.

\begin{figure}
	\centering
	\includegraphics[width=0.9\linewidth]{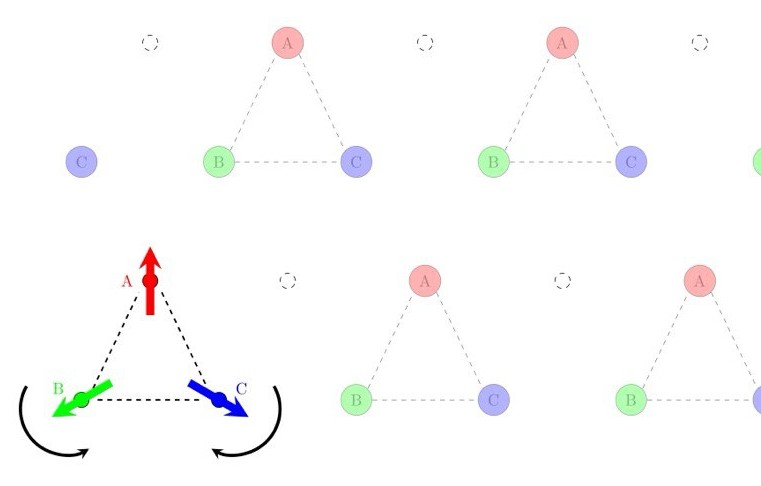}
	\caption{Illustration of the rotation of the moments B (counterclockwise) and C (clockwise) by the same angle $\alpha$.  Moment A remains fixed.}
	\label{fig:kagomerotation}
\end{figure}

\begin{figure}
	\centering
	\includegraphics[width=\linewidth]{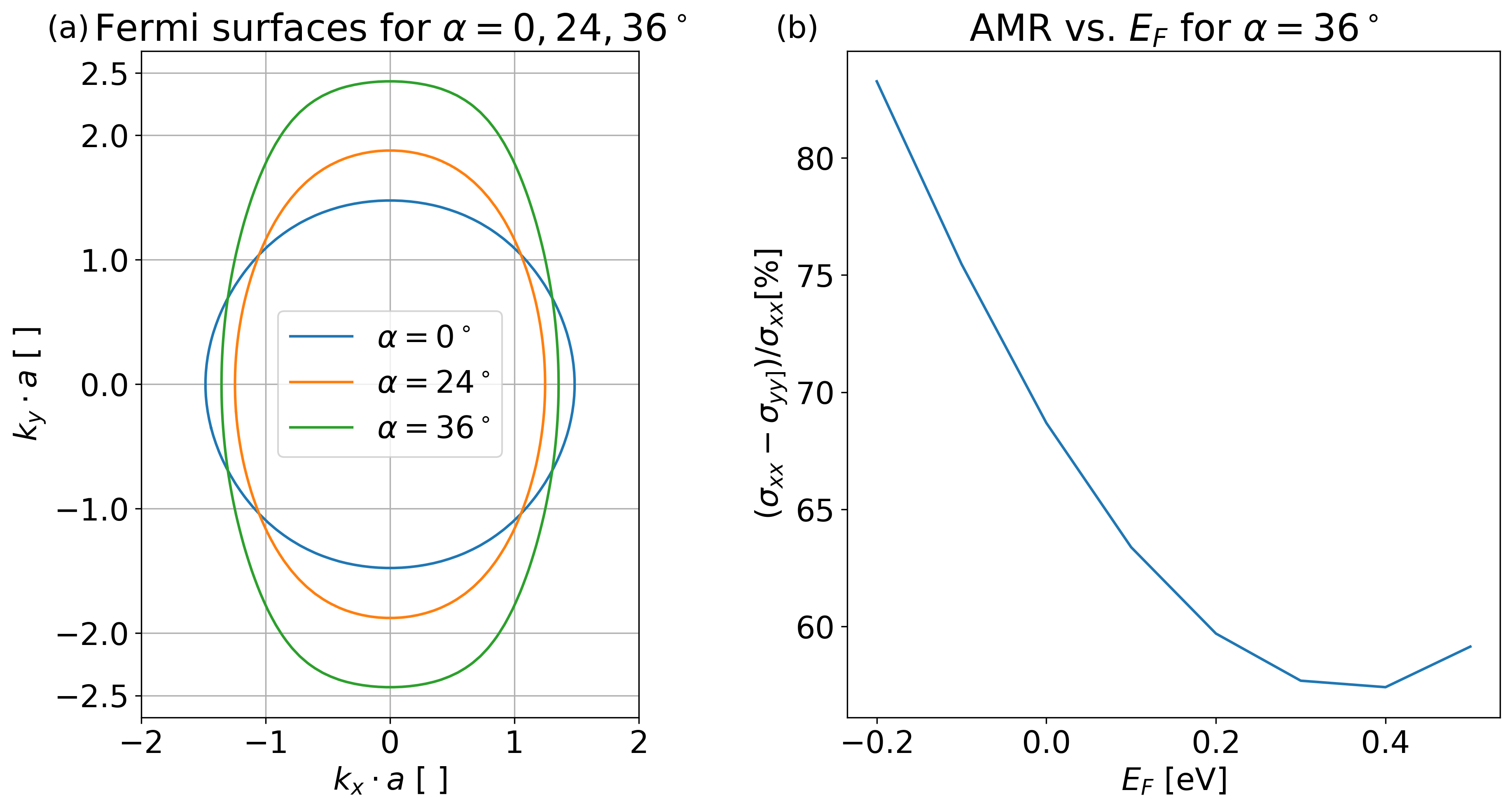}
	\caption{Anisotropy induced by magnetic order. (a) Fermi surface for the magnetic configuration corresponding to a rotation of moments B and C by~$\alpha = 0, 24, 36^\circ$, respectively, as defined in Fig.~\ref{fig:kagomerotation}. $E_F = 0$ ($\alpha = 0, 24^\circ$) and $E_F = 0.1$ ($\alpha = 36^\circ$). The partially compensated cases (PCM) $\alpha = 24^\circ$ and $36^\circ$ exhibit a pronounced and increasing anisotropy between $x$- and $y$-directions can be noted. (b) AMR vs. Fermi energy for $\alpha = 36^\circ$. The energy scale is directly comparable to Fig.~\ref{fig:totalkagomephase1a} (a)}
	\label{fig:asymmFS} 	
\end{figure}

To induce AMR in these systems, we consider instead a tilting of magnetic moments within the $xy$-plane, as illustrated in Fig.~\ref{fig:kagomerotation}: 
Here, moment $A$ remains fixed, while moments $B$ and $C$ are rotated---counterclockwise and clockwise, respectively---by the same angle $\alpha$.
This configuration qualitatively mimics the response to an external magnetic field or strain applied along the negative $y$-direction. Note that this simplified model neglects magnetocrystalline anisotropy and anisotropic exchange interactions. 
Several special cases are worth noting: $\alpha = 0^\circ (120^\circ)$ corresponds to the compensated states shown in Fig.~\ref{fig:totalkagomephase1a} (c) (Fig.~\ref{fig:totalkagomephase1a} (e)), $\alpha = 240^\circ$ corresponds to the ferromagnetic states with magnetization along the positive $y$-direction and $\alpha = 60^\circ$ corresponds to a collinear ferrimagnetic state. For all $\alpha \neq 0^\circ, 120^\circ, 240^\circ$ a partially compensated in-plane magnetization (PCM) exists. 
We find that these configurations lead to anisotropic Fermi surfaces (see Fig.~\ref{fig:asymmFS} for $\alpha = 24^\circ$ and $36^\circ$) and therefore exhibit intrinsic AMR. In contrast, the fully compensated ($\alpha = 0^\circ$) and ferromagnetic states yield isotropic Fermi surfaces and no AMR.

Applying the \textit{Symmetr} code~\cite{Symmetr} to get the generators of the real-space symmetries, we find that the PCM cases break both a $60^\circ$ and $120^\circ$ rotational symmetries around the $z$-axis, while the fully compensated and the FM case conserve at least one of them. 
The anisotropy in the collinear ferrimagnetic case ($\alpha = 60^\circ$) further demonstrates that the non-collinear magnetic order is not a prerequisite for AMR---low real-space symmetry alone is sufficient, as shown earlier in MnN.
The AMR in the PCM cases is not due to the existence of a net moment, but rather due to symmetry breaking, which can exist even in principle when a net moment is absent.

Turning to the triangular lattice (Fig.~\ref{fig:kagome_triangular} (b), we apply the same tilting procedure. However, in all cases—fully compensated, ferromagnetic, and partially compensated—we find no Fermi surface anisotropy and thus no intrinsic AMR. Symmetry analysis using \textit{Symmetr}~\cite{Symmetr} confirms that these configurations preserve either $60^\circ$ or $120^\circ$ rotational symmetry. 

In summary, we have analyzed a range of magnetic configurations on kagome and triangular lattices. Those that break the $60^\circ$ and $120^\circ$ real-space rotational symmetry around the $z$-axis exhibit anisotropic Fermi surfaces and thus AMR. These include both collinear and non-collinear ferrimagnetic configurations on the kagome lattice. In contrast, none of the triangular-lattice configurations investigated show intrinsic AMR.

\subsection{A Material Example: Mn$_3$Sn}
\label{sec_I_mat}

\begin{figure}
	\centering
	\includegraphics[width=1\linewidth]{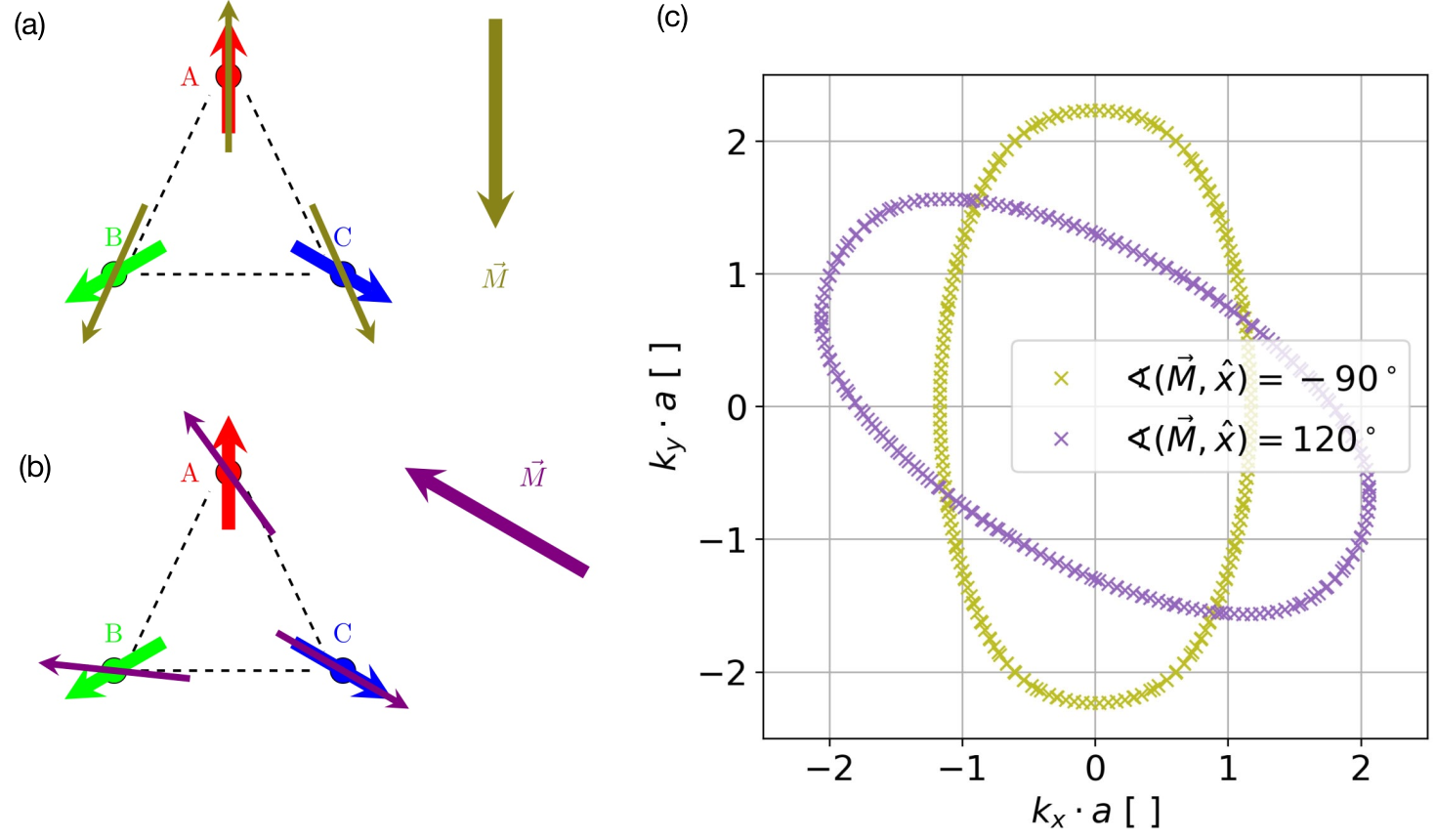}
	\caption{
    Illustration of symmetry breaking through in-plane tilting of magnetic moments in a kagome lattice (see Fig.~\ref{fig:kagome_triangular}(a)), leading to a small net magnetization $\vec{M}$ and modified Fermi surface.
	(a) The magnetic moments B and C were rotated by $\alpha = 36^\circ$ in negative $y$-direction (or $\sphericalangle (\mathbf{M}, \hat{x}) = -90^\circ$), causing a small magnetization in the same direction. (b) Simultaneously, the moments A and C were rotated by $\alpha = 36^\circ$ in $\sphericalangle (\mathbf{M}, \hat{x}) = 120^\circ$, causing a small magnetization in the same direction. Such tilts may arise, e.g., from external magnetic fields or strain. (Arrow lengths are schematic and not proportional to $|\mathbf{M}|$)
    (c) The resulting Fermi surfaces (ochre and violet, respectively) are rotated by 60$^\circ$ to each other, and do not overlap, leading to different AMR values.
        }
	\label{fig:fs-rotation}
\end{figure}

An example of a real material that can exhibit the phenomena discussed in the previous section is Mn$_3$Sn, a well-studied non-collinear antiferromagnet with a double-layer kagome lattice~\cite{Tomiyashi:1982}. 
Although known for over six decades~\cite{Cable:1993_a}, Mn$_3$Sn has recently attracted renewed interest due to its large anomalous Hall and Nernst effects, spin-polarized currents, and the presence of a tunneling magnetoresistance effect, making it a promising material for spintronic applications~\cite{Manna:2018, Chen:2021, Nakatsuji:2015, Zelezny:2017, Chen:2023}.
Using the \textit{Symmetr}~\cite{Symmetr} code, we confirm that the ideal magnetic configuration retains the real space rotations $60^\circ$ and $120^\circ$ around the $z$-axis, whereas tilted configurations (compare Fig.~\ref{fig:fs-rotation}) break both, consistent with our toy model results. 

The main effect of the magnetic field is a rigid rotation of all the moments~\cite{Tomiyashi:1982, Wu:2023}, however, magnetic field will also lead to tilting of the moments, which has been experimentally reported~\cite{Li:2021}. 
Furthermore, symmetry breaking may also arise from anisotropic bond distortions caused by piezomagnetism~\cite{Meng:2024} or hydrostatic pressure~\cite{Singh:2020}, both of which are sensitive to deviations from the ideal 3:1 stoichiometry. In off-stoichiometric samples, excess Mn may occupy interstitial or Sn sites~\cite{Gas:2025_a}, potentially contributing to magnetic disorder as suggested in Fig.~\ref{fig:kagome21}. We will examine scattering on such defects in the next section.
Figure~\ref{fig:fs-rotation} provides a conceptual illustration of how in-plane tilting of magnetic moments breaks the in-plane rotational symmetry and leads to AMR: The two configurations shown differ only in the relative orientation of the sublattice moments, resulting in rotated but non-overlapping Fermi surfaces, which lead to different values of AMR. This scenario could be experimentally probed by inducing symmetry-breaking tilts of the magnetic order, e.g., via external magnetic fields or strain.

In summary, Mn$_3$Sn provides a compelling platform for non-relativistic AMR. When the $60^\circ$ and $120^\circ$ symmetries are broken, anisotropic Fermi surfaces and AMR can emerge even without SOC. Measurements of transversal AMR (or PHE) by Sharma et al.~\cite{Sharma:2023_a} show non-saturating signal up to 9 T, which could be interpreted in accordance with Fig.~\ref{fig:asymmFS}(a): For a larger magnetic field, the tilt of the magnetic moments increases, leading to a larger anisotropy in the FS and thus a larger AMR.

\section{Extrinsic AMR}
\label{sec_extrinsic}

In this section, we investigate extrinsic (scattering-induced) AMR arising from aligned magnetic impurities (scenario sketched in Fig. 1b of Ref.~\onlinecite{Vyborny:2009}). 
The motivation comes from earlier studies on extrinsic AMR in dilute magnetic semiconductors \cite{Trushin:2009_a}. In those systems, one mechanism of AMR involves a SOC-induced spin texture at the Fermi level, which strongly resembles the non-relativistic spin texture found in non-collinear magnets, such as shown in Fig. \ref{fig:totalkagomephase1a}d,f. 
In the non-relativistic limit, such a texture cannot result in AMR by itself (considering an isotropic impurity) since it is not coupled to the lattice. However, AMR may arise when we introduce spin-dependent scattering. The basic idea is that electrons flowing in different directions carry different spins, which will then result in different resistance in the presence of spin-dependent scattering. 
We consider scattering from magnetic impurities, which we assume are only weakly coupled to the magnetic order and can be aligned by weak external magnetic field that does not perturb the magnetic order. This is the case of Mn impurities in antiferromagnetic salts\cite{Mischler:1977,Fuji:1968} or presumably in alloys\cite{Long:1993_a}
where they are bound only weakly via exchange interactions or are frustrated and exhibit nearly paramagnetic behavior. Similar observations of paramagnetic impurity behavior have been reported in topological insulator / ferromagnetic insulator heterostructures~\cite{Chiba:2017}. This mechanism of AMR closely resembles AMR in the dilute magnetic semiconductor (Ga,Mn)As,\cite{Vyborny:2009}, although in contrast to that case, the FM aligned impurities here change the symmetry of the system, and this is what enables the non-relativistic AMR.

\begin{figure}
	\centering
	\includegraphics[width=\linewidth]{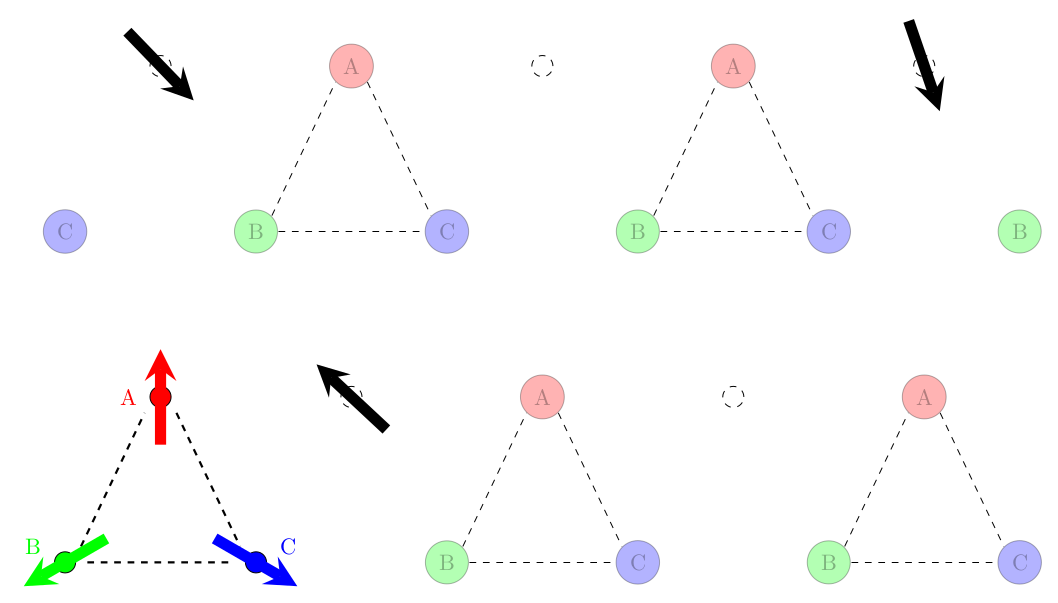}
	\caption{Illustration of unaligned magnetic impurities (black arrows) in the matrix of a kagome lattice. In Mn$_{3+x}$Sn$_{1-x}$ (where only a fraction of $x$ Sn atoms is replaced by additional Mn) this could correspond to substitutional impurities. The black arrows serve illustrative purposes as the concentration of impurities in realistic scenarios is expected to be significantly lower.
	}
	\label{fig:kagome21}
\end{figure}

We consider kagome and triangular lattices. In the kagome case, Fig.\ref{fig:kagome21} (black arrow) illustrates the effect of unaligned magnetic impurities. We assume these impurities are weakly coupled to the lattice and align under a magnetic field too weak to disturb the host magnetic order: As discussed earlier, such impurities may arise in Mn$_3$Sn due to off-stoichiometry, where excess Mn substitutes for Sn~\cite{Gas:2025_a}. While speculative, it is plausible that these Mn-rich regions act as magnetic impurities with similar behavior.

We will revert to the full treatment of scattering via Eq.~\ref{eq_FermiGoldenRule_1}, assuming magnetic impurities pointing in $\mathbf{i}$-direction described by the transition matrix $\hat{M} = \hat{S}_i \otimes \hat{1}_{N\times N}$, where $\hat{S}_i$ is the $i$-th Pauli spin matrix and $\hat{1}_{N\times N}$ is the $N$-dimensional identity matrix, where $N$ is the number of atoms in the unit cell. For simplicity, we will restrict ourselves to impurities either pointing in \textit{x}-direction,
which shall be abbreviated as \textit{x}-impurities, and impurities in \textit{y}-direction, denoted as \textit{y}-impurities. We calculate the conductivities $\sigma_{xx}$ and $\sigma_{yy}$ using the Eqs.~\ref{eq_Boltzmann_1}-\ref{eq_transmatrix}. The results are summarized in Table~\ref{T_extrinsic}.

\begin{table}
	\begin{tabular}{llll}
		Lattice & Impurity $\hat{M}$ & AMR? & Spin texture \\
		\hline 
		\textbf{Kagome (comp.)} & \textbf{\textit{x}} & \textbf{Yes} & \textbf{\textit{xy}} \\
		\textbf{Kagome (comp.)} & \textbf{\textit{y}} & \textbf{Yes} & \textbf{\textit{xy}} \\
		\textbf{Kagome (PCM)} & \textbf{\textit{x}} & \textbf{Yes} & \textbf{\textit{xy}} \\
		\textbf{Kagome (PCM)} & \textbf{\textit{y}} & \textbf{Yes} & \textbf{\textit{xy}} \\
		\hline 
		Triangular (comp.) & \textit{y} & extremal & \textit{z} \\
		Triangular (PCM) & \textit{y} & No & \textit{zy} \\
	\end{tabular}
	\caption{Summary of the extrinsic AMR calculations for kagome and triangular lattice in both compensated and partially compensated magnetic (PCM) configurations. The impurities as introduced in the text. "Yes" means that AMR was found, hence $\sigma_{xx} \neq \sigma_{yy}$ while for "No" an isotropic behavior was identified, where $\sigma_{xx} = \sigma_{yy}$. The value of the AMR ratio is not stated here, as our model is qualitative. "Extremal" means that both $\sigma_{xx}$ and $\sigma_{yy}$ are divergent in the absence of other scattering mechanisms (infinite in ideal case).
    "Spin texture" indicates the active components of the k-space spin texture at the FS. \textit{xy} indicates that all the spins are within the \textit{xy}-plane and thus $s_z = 0$ for all spins. The spin texture allows to develop intuition for whether a certain impurity would lead to suppressed scattering.}
	\label{T_extrinsic}
\end{table}

In the kagome case, both the \textit{x}- and the \textit{y}-impurity in either compensated and partially compensated (PCM non-collinear) cases cause non-zero extrinsic AMR. In the triangular case, however, no extrinsic AMR has been observed at all. In one specific case, the results in Tab.~\ref{T_extrinsic} are denoted by \textit{extremal}, which means that both $\sigma_{xx}$ and $\sigma_{yy}$ are infinite \textit{in our idealized model}. This originates from suppressed scattering due to $\Gamma \rightarrow 0$ for such impurity and the fact that the inverse scattering rate is part of the Boltzmann equation Eq.~\ref{eq_Boltzmann_1}. A similar situation occurred in the analysis of AMR in (Ga,Mn)As~\cite{Vyborny:2009}, where their minimal model yielded a singularity in the conductivity due to vanishing scattering matrix element. When including more terms to Eq.~9,  the original singularity transformed into a pronounced peak. Likewise, the divergence in our model is due to the neglect of any other scattering mechanism (e.g., all other types of impurities, phonons), and a real system is expected to display only a maximum rather than a singularity.

\begin{figure}
\includegraphics[scale=0.26]{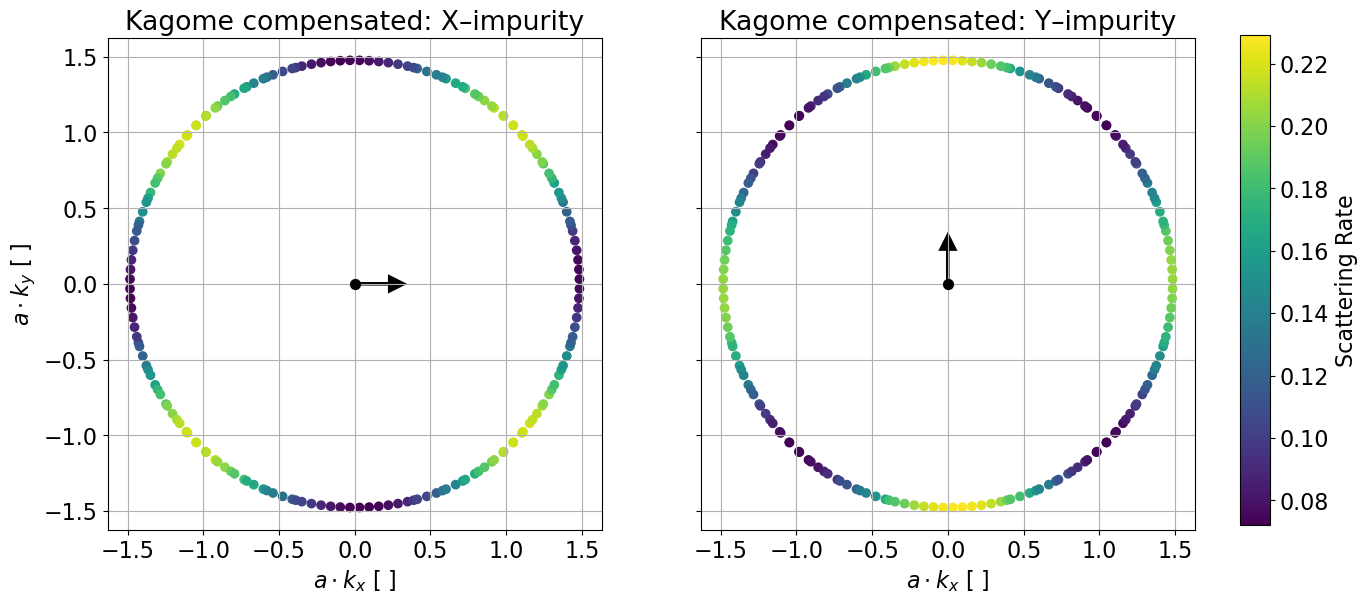}
\caption{Kagome lattice with magnetic impurity. Scattering rate (colour coded) on
the Fermi surface (see dashed line in Fig.~\ref{fig:totalkagomephase1a}a) for the same underlying
magnetic order (see Fig.~\ref{fig:kagome21}) and two orientations of the magnetic impurity. The left panel shows results for an $x$-impurity, while the right panel corresponds to a $y$-impurity. The impurity orientation modifies the angular dependence of the scattering, resulting in distinct conductivity responses and thus AMR}
\label{fig-09}
\end{figure}

In the last part of this section, we investigate the crystalline AMR, which refers to the anisotropy of conductivity created by the crystalline symmetry. Here, it can be defined, in analogy to Eqs.~\ref{AMR_Eq3} and \ref{AMR_Eq4}, as:

\begin{equation}
	\text{AMR}^{cry}_{vw} = \frac{\sigma_{vv} (\hat{M} = \hat{S}_v) - \sigma_{ww}(\hat{M} = \hat{S}_w)}{\sigma_{vv} (\hat{M} = \hat{S}_v)}
\end{equation}
where \textit{v} and \textit{w} are two arbitrary directions (e.g. \textit{x} and \textit{y}). The conductivity is measured along direction $v$ with impurities aligned in the same direction, and analogously for direction $w$. A non-zero value indicates an anisotropy arising from the crystal lattice itself.

The crystalline AMR is thus defined as the normalized difference of the longitudinal conductivity in \textit{v}-direction for a magnetic impurity in the same direction with the longitudinal conductivity in \textit{w}-direction for a magnetic impurity in the same direction. In both cases, the conductivity is measured along a given direction with the magnetic impurity aligned parallel to that same direction. The resulting anisotropy $\text{AMR}^{cry}_{vw} \neq 0$ would thus arise only from the influence of the crystal directions. 
For the kagome lattice, we indeed find non-zero crystalline AMR. This is illustrated in Fig.~\ref{fig-09},  where the Fermi surface and corresponding scattering rates are shown for $x$- and $y$-impurities.
The scattering rate of the $y$-impurity is not only rotated with respect to the $x$-impurity, but exhibits an additional two-fold modulation---scattering maxima are enhanced along the $y$-direction compared to the $x$-direction---resulting in different conductivities.

\section{Summary and Conclusions}
\label{sec_Sum}

In this paper, we investigated several mechanisms through which magnetic order can induce anisotropy in the conductivity tensor in absence of spin–orbit coupling, i.e., non-relativistic AMR. 
For this to occur, the magnetic order must break the real-space rotational symmetry of the crystal. In cubic MnN, the A-type AFM order breaks the $90^\circ$ rotational symmetry, giving rise to spontaneous AMR. On a kagome lattice, the relevant rotational symmmetries are $60^\circ$ and $120^\circ$, and breaking these---e.g., by manipulating the magnetic moments---leads to AMR. Our results suggest that non-relativistic AMR could be experimentally probed by inducing such symmetry-breaking tilts, for instance via magnetic fields or strain. We applied this theoretical framework to Mn$_3$Sn, a well-studied non-collinear AFM and spintronic application candidate. We confirmed that in this system, AMR emerges when the $60^\circ$ and $120^\circ$ rotational symmetries are broken, and this can occur without SOC. In the second part of this work, we investigated extrinsic AMR, which arises from spin-dependent scattering on magnetic impurities. We showed that even in magnetic configurations that are otherwise isotropic---such as the triangular order on the kagome lattice---AMR can appear if weakly coupled, aligned impurities are introduced. Specifically, we demonstrated that in a kagome lattice, such impurities can induce non-relativistic AMR, while in triangular lattices the effect is absent under similar conditions.

\section*{Acknowledgments}

We gratefully acknowledge insightful discussions with colleagues interested in anisotropic magnetoresistance (AMR). This work was supported by the Czech Science Foundation GA\v{C}R under Grant No. 22-21974S and TERAFIT project
No. CZ.02.01.01/00/22 008/0004594 funded by OP-JAK, call Excellent Research.
 
\begin{appendix}

\section{An Expanded Phenomenological Model}
\label{apx_phenomodel}

The angle-dependent form of AMR can be expressed phenomenologically in terms of the power expansion of the magnetization direction~\cite{Doring:1938,Limmer:2008,DeRanieri:2008}. 
This allows for the description of more complex crystalline AMR signals~\cite{Ritzinger:2021, Gonzalez-Betancourt:2024, NamHai:2012}. However, these models typically assume the presence of an SSA, such as the magnetization or N\'eel vector. In non-collinear systems, such an SSA does not exist---even in non-compensated cases induced by an applied magnetic field or strain---making it likely an oversimplification to ignore the role of the MSLs. The idea of a "local" treatment is not new. For instance, basic AMR models in FMs rely on separate contributions for spin-up and spin-down electrons (two-current models)~\cite{Ritzinger:2023}, and the Edelstein effect in non-collinear Mn$_3$Sn can be calculated on a sublattice-resolved basis~\cite{Gonzalez-Hernandez:2024}. A similar approach can be applied to AMR by treating the each MSL individually within the phenomenological model, yielding:

\begin{equation}
	\rho_{yy} = \rho_0 + \sum_{m = 1,2,3} \sum_{n = 2, 4, 6, ...} c_{m,n} \cos(n \alpha_m)
	\label{eq_sublattice_AMR}
\end{equation}
where $m$ is the index of the MSL, $n$ is the order of the spherical harmonic, $c_{m,n}$ is the index of the $n$-th harmonic of the $m$-th MSL, and $\alpha_m$ is the angle of the magnetization direction of the $m$-th magnetic moment (assuming an in-plane rotation). The coefficients $c_{m,n}$ can be obtained by fitting to experimental data, although doing so for multiple MSLs requires measurements at different magnetic field strengths. The angular positions $\alpha_m$ of the magnetic moments can be obtained using a Stoner-Wohlfahrth (SW) model. While not the main focus of this work, Eq.~\ref{eq_sublattice_AMR} combined with a suitable SW model could help to disentangle MCA from AMR in non-collinear systems.

\end{appendix}

\def\urlprefix{}
\def\url#1{}

\end{document}